\documentclass[journal]{IEEEtran}
\usepackage{cite}
\usepackage{amsmath}
\usepackage{tikz}
\usepackage{comment}
\usepackage{graphicx}
\usepackage{caption}
\usepackage{subcaption}
\usepackage[english]{babel}
\newtheorem{theorem}{Theorem}
\newtheorem{remark}{Remark}
\newtheorem{corollary}{Corollary}
\newtheorem{lemma}[theorem]{Lemma}

\ifCLASSINFOpdf
\else
\fi
\begin{document}
\title{On the Dynamics of Control}
\author{
    \IEEEauthorblockN{Rachit Mehra\IEEEauthorrefmark{1}, M Parimi \IEEEauthorrefmark{2}, S.R. Wagh\IEEEauthorrefmark{3}, Navdeep M Singh\IEEEauthorrefmark{4}}\\
    \IEEEauthorblockA{\IEEEauthorrefmark{1}TenneT Offshore GmBH
    \{rachit.mehra\}@tennet.eu}\\
    \IEEEauthorblockA{\IEEEauthorrefmark{2}Research Scholar, $EMC^2$ Lab,Veermata Jijabai Technological Institute (VJTI)}\\
    \IEEEauthorblockA{\IEEEauthorrefmark{3}Assistant Professor, Electrical Engineering Department (EED), VJTI}\\
    \IEEEauthorblockA{\IEEEauthorrefmark{4}Chair Professor, $EMC^2$ Lab, VJTI, Mumbai, India}

\thanks{ }
}
\maketitle
\begin{abstract}
We present a dynamical system approach for the control of a nonlinear dynamical system by defining the control problem in a Fiber bundle framework. The constructive procedure derived results in the generation of a NHIM/NAIM which facilitates the use of tools and ideas from dynamical system theory to analyze and understand the properties associated with the controlled system. The time scale separation, decoupling of system dynamics, and their role in the system behavior are analyzed. An overview of the benefits of the above approach is demonstrated by briefly discussing three main application areas.
\end{abstract}

\IEEEpeerreviewmaketitle

\section{\label{chap:intro}Introduction}%
Control synthesis for general nonlinear systems remains a challenging problem and no one technique is universally acceptable. Two popular classes of solutions are explicit control constructions based on Classical Lyapunov theory and model predictive techniques involving real time optimization. 

A Lyapunov function characterizes the stability of a system and is related to the intuitive idea of energy decaying in stable systems. For this reason a Lyapunov function is a quadratic function or convex in nature and is a dominant idea in control theory, both in analysis and design.

On the design side, the Controlled Lyapunov function (CLF) provides a necessary and sufficient condition for controllability of a large class of systems, especially for systems affine in control, makes the construction of a controller a feasible problem (given a CLF) \cite{Artstein}, \cite{SONTAG1989117}, \cite{Freeman}. However, in general, it is difficult to find a CLF. Thus, constructive control design for nonlinear systems remains a challenging problem even in the case of full state feedback \cite{Slotine1991}, \cite{Isidori}, \cite{Kokotovic}. Most of the present constructive design tools tend to be limited in their scope of applications. Backstepping and related methods provide a systematic and constructive procedure \cite{Krstic} but are generally limited to systems of a particular triangular structure. For the control of underactuated mechanical systems, energy and passivity based methods are the natural choice, given that the idea of energy (Kinetic and Potential) is well defined and forms the basis of modeling systems either in the Lagrangian form or the Hamiltonian framework \cite{Bloch}, \cite{Ortega2002}, \cite{Mehra2017}.

An alternative to the above explicit design approach has been the development of the Control Contraction Metric (CCM) approach and the Model Predictive Control (MPC) approach to controller design. Both of these methods are dependent on certain numerical computations and optimization procedure performed either offline or online. This imposes computational burden and could lead to numerical instability especially when handling highly nonlinear systems.
\cite{Diehl2009}. In the case of MPC it generally remains difficult to predict or analyze the performance of non-linear MPC schemes by any method other than exhaustive simulations. On the other side of the spectrum of the above control methods lie the variable structure methods, which have gained some fraction, but issues related to their inherent discontinuous behavior limit their application \cite{Shtessel2014}.

In this write-up, we address the challenges associated with control design of both linear and nonlinear systems by viewing the system as a dynamical system modeled by an ordinary differential equation and by bringing to bear the ideas from dynamical system to facilitate the control design. This method provides better insight into the behavior of the system, and thus a wide spectrum of issues related to design and control can be addressed. A constructive approach is developed to compute the normally hyperbolic invariant manifold (NHIM) by considering a continuation of the Passivity and Immersion (P\&I) method. 
The proposed methodology has been applied to wide range of application areas e.g. constructive approach has been proposed to obtain Control Lyapunov Functions (CLF) aimed for stabilization and control of nonlinear systems, cascade control laws have been derived by achieving time scale separation between the subsystems, and lastly for the problem of parameter estimation better performance is achieved through the decoupling of repressor dynamics.

Recent advances in the area of stabilization of underactuated mechanical systems includes \cite{Mehra2017}, \cite{nayyer2022towards} where the geometric approach to constructive control methodology was proposed. The idea of fibre bundles, splitting, and generation of passive output was adapted to nonlinear state space formulation through the target dynamics and its associated manifold in \cite{NayyerGauss2022}, \cite{NayyerCSL2023}. The methodology was extended to analyze and design optimization and parameter estimation techniques in \cite{GunjalCSL2024}, \cite{gunjal2024unified}.

The remainder of this paper is organized as follows. In Section 2 we extend the P\&I method  detailing the constructive steps. Section 3 extends the proposed approach to compute NHIM. The three application areas and related examples are illustrated in Section 4 and the conclusion is outlined in Section 5.%
\section{\label{chap:meth}Passivity and Immersion (P\&I) Approach}%

We start by posing the control problem in the appropriate mathematical framework to facilitate the deployment of tools from dynamical system theory.
The concept of the fiber bundle and its associated Ehresmann connection (EC), which happens to be a key concept needed to describe the geometry associated with a given differential manifold, are introduced. The concept of EC is one of the most general definitions of connections in differential geometry and it applies to any smooth fiber bundle thus making it more versatile than the other types of Connections that may have other structural requirements. 
For example, unlike affine connections that focus on parallel transport of vectors,  the EC defines a direct sum decomposition of the tangent space into a horizontal subspace and a vertical subspace in the total space of the fiber bundle. This allows for a broader mathematical framework that can be applied to various geometrical structures.

The notation, basic definitions, and representation of control system in fiber bundle structure are introduced in the Appendix. The local representation of EC and the integrability of connection are detailed, thus preparing the framework to propose the methodology.


%

The control approach to be adopted is highly dependent on the nature of the connection. 
\paragraph{Flat Connection} In case of a flat connection, the Ehresmann connection $\nabla \phi(x)$ is one form and is required to be exact (integrable). This ensures that the horizontal spaces (implicit manifold) defined by $ V_H$ are Frobenius integrable, allowing transverse sections of the fiber bundles. This implies that the curvature vanishes identically, implying that the connection is flat, thus ensuring parallel transport of vectors along curves without twisting. The direct sum decomposition of the tangent bundle $TM$ allows for smooth feedback control and the proposed P\&I method  builds upon this to obtain the appropriate feedback control laws.

\paragraph{Non flat Connection} If a connection is not flat, there are topological obstructions present. The implicit manifold is nonexistent, and thus finding a feasible smooth feedback control law to stabilize the system is not possible. The non-existence of the implicit manifold (and smooth feedback stabilizing control law) may not be directly related to the nature of the control system's vector field, but in the case of non-holonomic systems the non-integrability of the system vector field has a direct bearing on the non existence of implicit manifold and thus a continuous stabilizing feedback law does not exist \cite{Moulay2023}. In such case, discontinuous feedback or time-varying control is applicable.
\begin{remark}
The smooth flow on an integrable manifold $P$ can lie on an immersed manifold $M$ of $P$, iff $M$ is integrable. The non-integrability of a manifold $M$ implies that the tangent spaces of $M$ do not form a consistent distribution that supports smooth flows globally.    
\end{remark}
 
\begin{figure}[h!]
    \centering
    \includegraphics[scale=0.8]{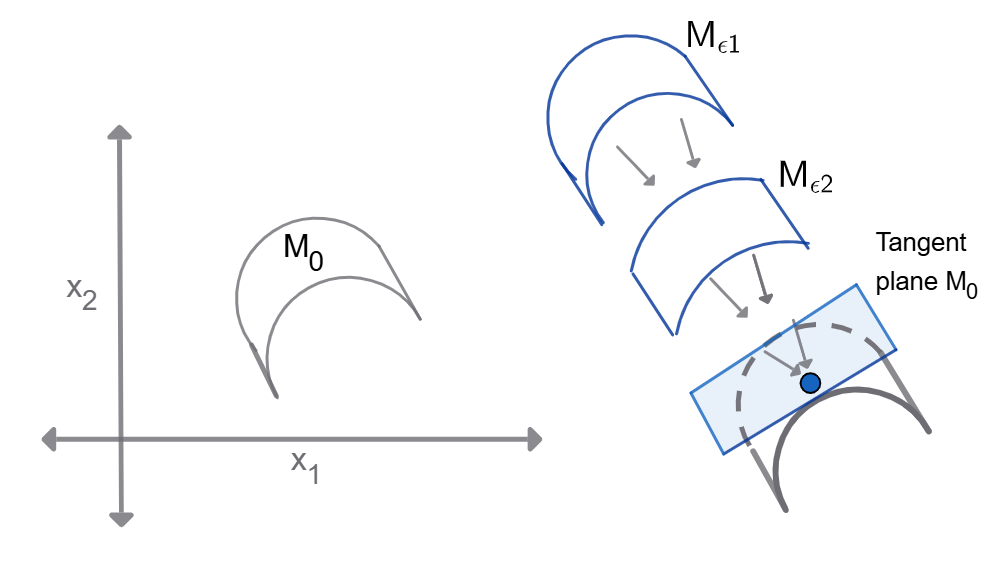}
    \vspace{0.2cm}
    \caption{Invariant manifold $M_0$ and the effect of perturbation on $M_0$}
    \label{fig:sp}
\end{figure}

In this write-up, we will restrict ourselves to the case of a flat connection. 


Consider the control system
\begin{equation}\label{eq:oe}
 \dot{\mathrm{x}} = f({\mathrm{x},\lambda}): ({\mathrm{x
}},\lambda) \in X, {\mathrm{u} \in R^{n}} \end{equation}
\begin{equation}\label{eq:pi2}
\dot{\mathrm{\lambda}} = g({\mathrm{x},\lambda})  + u : ({\mathrm{x
}},\lambda) \in M\end{equation} 

where $X = (x, \lambda) \in R^{n-1} X R^1$ is a $n$ dimensional state space. 

Based on the control objective, we define the target dynamics, $$\lambda = \phi (x): \dot {\mathrm {x}}=\mathrm {f(\mathrm {x}, \phi(x))}$$
i.e. $u$ is to be determined such that all system trajectories reach the target manifold (implicit manifold). 
\begin{equation}\label{eq:IM}
M_0(x,\lambda) = \{{({\mathrm{x}},\lambda) | \lambda + \phi(x) = 0}\} : (x^*, \lambda^*) \in M_0     
\end{equation}

The steps to obtain the control law are detailed as follows :
\begin{enumerate}
    \item The normal to the implicit manifold $M$ is defined as follows: 
$$\nabla M_0 = \begin{bmatrix}
\nabla \phi(x) & 1 
\end{bmatrix}$$
    \item Define a symmetric degenerate rank one matrix $R$ as 
    $$ R = \nabla M_0^T \nabla M_0= \begin{bmatrix}
\nabla \phi^T \nabla \phi & \nabla \phi \\
\nabla \phi & 1 
\end{bmatrix}$$
We label $R$ as the semi-Riemannian metric. 
\item Using the fiber bundle structure of the control system and $R$ as defined above, we invoke the idea of the Ehresmann connection on the fiber bundle. 
$$ R = \nabla M_0^T \nabla M_0= \begin{bmatrix}
\nabla \phi^T \nabla \phi & \nabla \phi \\
\nabla \phi & 1 \end{bmatrix} = \begin{bmatrix}
m_{11} & m_{12} \\
m_{21} & m_{22} 
\end{bmatrix}$$


\item The connection term $\nabla \phi = m_{21}m_{22}^{-1}$ leads to the direct sum decomposition of the tangent bundle $TM$ as follows;
\begin{equation} \nonumber
    TM=V_H\oplus V_V=(\dot x  \hspace{0.2cm} -\nabla \phi \dot x) \oplus (0 \hspace{0.2cm} \dot \lambda+ \nabla \phi \dot x)
\end{equation}
i.e. $\forall p \in M$,  the tangent space $T_pM$ at $p$ decomposes as:
$$T_{p}M = V_{H_{p}} \oplus V_{V_{p}}=(\dot x  \hspace{0.2cm} -\nabla \phi \dot x)_p \oplus (0 \hspace{0.2cm} \dot \lambda+ \nabla \phi \dot x)_p$$
for each $(\dot x , \dot \lambda)_p \in T_p M $.


\item As $u$ acts along $V_V$ the passive output $y$ is chosen as $y= \dot{\lambda} + m_{21}m_{22}^{-1} \dot{x}$.  If the passive output is integrable i.e. $m_{21}m_{22}^{-1} \dot{x} = \nabla \phi \dot{x}$ is integrable, implying that $\phi(x)$ is an integrable connection \cite{nayyer2022towards} then one can define the storage function as: 
$$S = \frac{1}{2} \left( \int y  \hspace{0.1cm} dt \right)^2$$

For an integrable connection $\phi(x)$  the storage function can also be represented as 
\begin{equation}\label{eq_S}
S = \frac{1}{2} \left( \int y  \hspace{0.1cm} dt \right)^2 = \frac{1}{2} M^2  
\end{equation}
where 
\begin{equation} \nonumber
M(x,\lambda) = \{{({\mathrm{x}},\lambda) | \lambda + \phi(x) \neq 0}\}   
\end{equation}
where $M$ is defined for $(x, \lambda)$ not lying on $M_0$
\item The control law $u$ follows from the condition that 
$$\dot{S} \leq - \hat{\alpha} S $$
From (\ref{eq_S}) it follows that
$$M \dot{M} \leq \frac{\hat{\alpha}}{2} M^2M$$ \\ 
i.e. $\dot{M} \leq - \alpha M $, where $\alpha = \frac{\hat{\alpha}}{2}$ \\

Now choosing $u$ such that the $\dot{M} \leq \alpha M$, leads to 
\begin{equation}\label{eq_CL}
    u = - \nabla\phi (x) \dot{x} - \alpha M
\end{equation}
Thus the control system takes the form
 \begin{align}\label{eq:cs1}
 \dot{\mathrm{x}}& = f({\mathrm{x},\lambda}) \\
 \dot{\mathrm{\lambda}} &= - \nabla\phi(x)\dot x - \alpha M \label{eq:cs2}  
 \end{align}
Since the control law follows from the fact that $\dot{M} \leq \alpha M$ it follows that $M \rightarrow TM_{0(\dot{x}, \dot{\lambda})}$, where $(\dot{x}, \dot{\lambda})$ are the equilibrium points belonging to $M_0$.
An alternative way to look at \eqref{eq:cs2} is as follows:
$$ \dot{\mathrm{\lambda}} + \nabla\phi(x)\dot x =  - \alpha M$$

That is, the vector field or the direction along which the control law acts is along $V_V$ as is evident from the above equation. Thus, the controlled trajectories are along $V_V$ and orthogonal to $V_H$. We call the direction along $V_V$ the pathway defined by the Connection term $\nabla \phi (x) = m_{21}m_{22}^{-1}$ and this is the path along which the off manifold trajectories (of $M_0$)  approach $M_0$ under the control action $u$.

Finally, the system (\ref{eq:cs1} - \ref{eq:cs2}) can be interpreted as follows.
 \begin{align}\nonumber
 \dot{\mathrm{x}}& = f({\mathrm{x},\lambda}) \\
 \dot{\mathrm{\lambda}} &= \underbrace{- \nabla\phi(x)\dot x}_{Pathway} - \underbrace{\alpha M}_{Push-term}  \nonumber
 \end{align}
\begin{enumerate}
    \item The term $-\nabla \phi \dot x$ defines the geometric pathway for the controlled trajectories  (defined by the connection $\nabla \phi (x)$ and  along the $V_V$ direction as defined above)
    \item The second term $-\alpha M$ pushes the off manifold trajectories (not on $M_0$) to $M_0$ thus ensuring the attractivity of the manifold $M_0$
\end{enumerate}
\end{enumerate}
Summarizing: The system trajectories converge to the manifold $M_0$ at the exponential rate defined by the control Lyapunov function (CLF), $\dot S \leq -\alpha S$.\\

\begin{remark}
In the above derivations the signs associated with the pathway term in the control law depend on how the implicit manifold $M_0$ is defined. In the case where $M_0$ is defined as $M_0 =\left\{(x,y):\lambda-\nabla\phi(x)\neq0 \right\}$ then the sign in the control law changes and the equation (\eqref{eq:cs2}) is modified as $$\dot{\mathrm{\lambda}} = \nabla\phi(x)\dot x - \alpha M$$ 
\end{remark}

 
\begin{remark}
The EC defines a geometric pathway in the state space $M$, for the action of $u$. The target manifold is aligned with $V_H$ and the pathway for the control action is aligned along $V_V$ where $V_H\cap V_V={0}$. Control design involves steering the system from an initial state to a desired state (e.g. stabilizing and equilibrium point or tracking a trajectory). Natural geometric pathways in state space offer efficient routes for this purpose because they align with the system's inherent tendencies, reducing interactions in the system dynamics and the effort for control.
\end{remark}

Discussions : 
\begin{enumerate}
    \item The target manifold which contains the invariant set is attractive under the control action of $u$, i.e. the off-manifold trajectory converges to the manifold exponentially.
    \item The target manifold contains the desired invariant manifold which could be an equilibrium point or a periodic orbit, for example.
    \item If the target manifold is invariant (i.e. the target dynamics is exponentially stable as discussed in the example below), then the Fenichel's theorem is directly applicable and it persists under small perturbations.
    \item Thus target manifold helps to set up the geometrical background and provides the framework for the dynamical system theory of the Normally Hyperbolic Invariant Manifold (NMIM) /Normally Attractive Invariant Manifold (NAIM) and their various properties and applications in system theory
    \item In case of contraction, the equilibrium point is viewed as a NAIM
        \end{enumerate}
    
\section{Normally Hyperbolic Invariant manifold}
\textbf{Definition:} Let $M$ be a smooth compact sub-manifold (to be considered as implicit manifold) embedded in $ R^n$, with a dynamical system described by flow $f: X\rightarrow X$. The manifold $M$ is normally hyperbolic at a point $ p\in M$ if the tangent space of the ambient manifold at $ p\in M$ denoted by $T_pM$ admits a splitting that is invariant  differential of the dynamics. To introduce the hyperbolicity condition we first redefine the tangent bundle splitting as follows. 
\begin{enumerate}
    \item Tangent Bundle splitting: $$ T_pX=T_pM \oplus E_p^s\oplus E_p^u$$
where $E_p^s:$ denotes a stable bundle, corresponding to directions that contract under dynamics (stable directions)
$E_p^u:$ denotes the unstable bundle corresponding to directions that expand under the dynamics.
    \item Invariance: The splitting is invariant under the differential:
    $ Df_p:T_pX\rightarrow T_{f(p)} X$
    \item Hyperbolicity condition: The dynamics in the normal directions ($E_p^s,E_p^u $) dominate the dynamics in the tangential directions.
    \item Global Definition: The manifold $M$ is normally hyperbolic if the above splitting and (among other conditions) hyperbolicity holds for any $p \in M$. The tangent bundle $TM,E_p^s,$ and $E_p^u$ together form a Whitney sum over M: 
    $ TX/M=TM \oplus E^s \oplus E^u$.\\
\end{enumerate}

Considering the target dynamics, the above statement holds true if the target dynamics lies tangent to the manifold $M$ and NHIM $M$. However, when the above condition does not hold and the target dynamics is asymptotically stable, then the manifold $M$ qualifies as a normally hyperbolic manifold containing the invariant set that is the equilibrium point, which is viewed as being perturbed by the target dynamics. We have sufficient tools to state the following Lemma. 

\begin{lemma}
The proposed P\&I based control law \eqref{eq_CL} ensures the GAS/GES convergence of the original system \eqref{eq:oe} to invariant manifold $M$ \eqref{eq:IM} where the modified system dynamics, given by \eqref{eq:cs1}-\eqref{eq:cs2}, is NHIM under the control action. 
\end{lemma} 

Fenichel's theorem: The essence of the theorem is as follows. Suppose $S$ is a compact normally hyperbolic sub manifold (equilibrium point) of the implicit manifold $M$, then for $\epsilon>0$ sufficiently small, there exists a locally invariant manifold $S_\epsilon$ diffeomorphic to $S$. Roughly speaking local invariance implies that the trajectories can enter or leave $S_\epsilon$ only through its boundaries. \\
Note: If the implicit manifold is invariant, then by the application of the above theorem, it implies that $M$ is perturbed to a locally invariant manifold $M_\epsilon$ diffeomorphic to M.


\section{\label{chap:res}Examples}%
This section illustrates a few of the application areas, through examples, of the P\&I approach. The examples highlight the dynamical system approach to control design through the use of the NHIM/NAIM framework and the idea of how time scale separation of the controlled system dynamics can be achieved through the use of the P\&I approach. The examples are broadly divided as follows. In the first Subsection-A,  the constructive approach to control design is addressed based on the P\&I method. The second Subsection-B, focuses on time scale separation of the controlled system dynamics which results in improved transient response of the system. The parameter estimation problem for adaptive control systems is briefly discussed in Subsection C.
\subsection*{Example A: Control of Nonlinear/Linear systems}
Control Lyapunov functions (CLF) provide the theoretical bases for designing control laws and assessing the stability of control-affine systems. Despite the widespread applicability of CLF as a tool for nonlinear control design, finding a suitable Lyapunov function candidate that satisfies the desired objectives has always been challenging and non-trivial. The work \cite{astolfi2003immersion} attempts to develop controllers for nonlinear systems without the requirement of a CLF in the control law design phase. However, the lack of a rigorous procedure for determining the attractivity of the manifold limits the practical application of this approach. The proposed P\&I approach provides a constructive and systematic procedure to achieve stabilization and control of nonlinear systems. The approach is illustrated through examples of systems that are linear/nonlinear with the objective of finding a suitable control law to stabilize the system. 
\subsubsection*{A.1. Example 1}
Given a dynamical system:
$$ \dot x_{1}=x_1+x_2 $$ 
$$\dot x_2=u$$
The steps involved in applying the P\&I approach to find the control law $u$ are listed below:
\begin{enumerate}
    \item The target dynamics is chosen as: $\dot x_1=-x_1$ \\and the implicit manifold M defined as:\\
\indent $ M= \{ (x_1,x_2) \in R^2, x_2+\phi(x_1)=0 \}$ with \\ $\phi(x_1)=2x_1, \nabla \phi=2$
\item The control law is derived as $u=-\nabla \phi \dot x_1-\alpha M$
\item The controlled system dynamics are expressed as:\\
$ \dot x_{1}=x_1+x_2 $\\
$\dot x_2=-2(x_1+x_2)-(x_2+2x_1)$ (for $\hspace{0.1cm}\alpha=1)$\\
$=-4x_1-3x_2$
\item The above system of equations have an equilibrium point $(-1,-1)$ which is exponentially stable, with both the off-manifold dynamics and the target dynamics converging at the rate $-1$.
\item In general, if the target dynamics is exponentially stable (verified in most cases by using the Krasovkii lemma) then along with the (exponential) attractivity of the implicit manifold under the control action $u$, one obtains the exponential stability of the equilibrium. Further, if the rate of convergence is the same (of the target dynamics and the off manifold controlled dynamics), then the controlled system dynamics is said to be    contracting . If the attractivity of the manifold $M$ only is established, then one has horizontal contraction of the controlled system dynamics. 
\item To verify the above for $ S=\frac{1}{2}M^2=\frac{1}{2}(x_2+2x_1)^2$ \\If $ \dot S\leq -\alpha S $, then the manifold M is attractive.
$ \implies  (x_2+2x_1)(\dot x_2+2\dot x_1)\leq -\alpha S$\\
On simplifying, $-3(x_2+2x_1)^2\leq-\alpha S$ and if $\alpha=6$, then  attractivity of $M$ implies that the resultant dynamics is horizontally contracting (the nature of 
 target dynamics is not considered). The horizontal contraction under control action combined with exponentially converging target dynamics (verified through the application of the Krasovoskii lemma and Gronwall inequality) leads to contraction.
\end{enumerate}
\subsubsection*{A.2. Example 2}
Consider 
$$\dot x_1=x_2$$
$$\dot x_2=-13x_1+4x_2+u$$
The above equations represent highly oscillatory and unstable dynamics. The control $u$ is designed as follows:
\begin{enumerate}
    \item Let the target dynamics be: $\dot x_1=-x_1$, one has the implicit manifold $M=\{(x_1,x_2)\in R^2|x_2+x_1=0 \}$.
    \item $\phi(x)= x_1, \nabla \phi=1$
    \item This leads to: $ u=-\nabla \phi_{x_1} \dot x_1-\alpha M=-x_1-2x_2$ (with $\alpha=1$) 
    \item The resultant dynamics are: 
 $$\dot x_1=x_2$$
$$\dot x_2=-x_1-2x_2$$ with roots being at (-1,-1).
\item We thus have a NAIM where the slow invariant manifold is $ x_1+x_2=0$  with tangential dynamics $\dot x_1=-x_1$ and the fast transverse dynamics being $\dot x_2=-x_1-2x_2$.
\end{enumerate}

\subsubsection*{A.3. Example 3} Given the system below (Example 6 \cite{astolfi2003immersion}) : 
$$\dot x_1=-x_1+x_1^3x_2$$
$$\dot x_2=u$$ 
It is desired to find a suitable $u$ to stabilize the origin (0,0). Using the P\&I approach:
\begin{enumerate}
    \item Let the target dynamics be of the form: $\dot x_1=f(x_1,\phi(x_1))$
    \item  Choose the manifold $M=\{(x_1,x_2)\in R^2 \mid x_2=-x_1^2\}$, i.e  $M: \{x_2+x_1^2=0  (or) 
 x_2+\phi(x_1)=0) \}$
    \item Compute $\nabla \phi(x)=2x_1$
    \item The control law which steers the off-manifold trajectories to M is of the form:\\ $ u=-\nabla \phi \dot x_1-\alpha(x_2+x_1^2) =-\nabla \phi \dot x_1-(x_2+x_1^2)$ ( for $\alpha=1$).
 \item The controlled system dynamics is then of the form:
 \begin{align}\label{eq:a11}
 \dot x_1=-x_1+x_1^3x_2\nonumber \\
\dot x_2=x_1^2-2x_1^4x_2-x_2
\end{align}
\item The control law $u$ ensures that off-manifold trajectories converge to the target manifold exponentially (owing to the splitting of the tangent space into $V_H$ and $V_V$) .
\item The target dynamics converges asymptotically to the equilibrium point, which can be verified using the Krasovskii's lemma. Defining a storage function $V$ as:
 $$V(\dot x_1)=\frac{1}{2}\dot x_1^2$$
$$\dot V(\dot x_1)=\dot x_1\ddot x_1= \dot x_1 (-\dot x_1-5x_1^4\dot x_1)=-\dot x_1(1+5x_1^4) \dot x_1$$
\item This implies that $\dot V_1 (\dot x_1)\leq0$, i.e. the target dynamics converge asymptotically to the equilibrium point $(0,0)$. Hence, the target manifold is not invariant for the given target dynamics. The dynamics transverse to the target manifold convergence exponentially (as they are steered by $u$ along $V_V$ whose direction is conjugate to $ V_H$) but the dynamics along the manifold converges asymptotically hence the overall system dynamics cannot be contracting. As a result, the equilibrium is asymptotically stable.
\end{enumerate}
\begin{remark}
In \cite{astolfi2003immersion}, the same control law as derived above (\ref{eq:a11}) has been obtained by a method that results in some lengthy calculations. As reported in \cite{astolfi2003immersion} the method was formulated to overcome one of the serious limitations of the I\&I method, i.e. the absence of any constructive method to prove the attractivity of the target manifold.    
\end{remark}

\indent In general terms, the method uses the Lyapunov-Finsler theory and proceeds to prove (horizontal contraction) the attractivity of the target manifold \cite{wang2016immersion}.\\  \indent Using the variational dynamics of the system and semi-Riemann metric defining the geometry of the target manifold a variational form $\delta u$ of the desired control $u$ is obtained. To go from $\delta u$  to $u$ one needs to integrate along the geodesic between two adjacent converging trajectories. In the case of flat geometry (applicable for the above example) the process of integration is along the line segment joining the two trajectories. The approach of \cite{wang2015immersion} seems to be inspired in part by the Control Contraction Metric (CCM) approach, i.e. the variational formulation and path integration method as employed in the CCM. (but for the integration along the line segment employed). However, in general, for non-flat or general Riemannian manifold (CCM) one needs to calculate the geodesic between nearby or two adjacent trajectories so that $u$ can be calculated by integration along the geodesic. Finding the desired geodesics along which the form $\delta u$ can be integrated further adds to the various steps that are needed to derive the control law. The advantage of the geometrical perspective associated with the P\&I control approach allows one to obtain the same control law as derived in (\ref{eq:a11}) in a straightforward manner and in a few short steps.
\subsection*{Example B: Time-Scale Separation}
This example section illustrates the scope of the P\&I approach to applications that involve the the control of systems with different time scales and also the cascade control architecture. The objective is to achieve a time-scale separation between the fast and slow dynamics of a system of coupled equations. This separation yields an improvement in the transient response of the system.\\ 
Consider the interconnection of two systems described by vector fields $f_{i}(.)$ on $x_{i} \in R^n $ respectively for $i=1,2$.
\begin{align}\label{eq:b21}
\begin{gathered}
\dot x_{1} = f_{1}(x_{1},x_{2}) \\
\dot x_{2} = f_{2}(x_{1},x_{2})
\end{gathered}
\end{align}
\indent The study of such interconnections is, in general, challenging. One way is to assume that the two systems above evolve on separate time scales. \\
If $x_{2}$ evolves at a faster time scale  as compared to $x_{1}$,then on a fast time scale the dynamics of $x_{2} $ are replaced by the algebraic equation $f_{2}(x_{1},x_{2})= 0$.\\
This two-time-scale interconnection is represented by a differential algebraic equation system $S_{1}$ and a boundary layer system $S_{2}$:
\begin{align}\label{eq:b22}
\begin{gathered}
S_{1} : \dot x_{1} = f_{1}(x_{1},x_{2}) s.t. f_{2}(x_{1},x_{2}) = 0  \\
S_{2} : \frac{dx_{2}}{dt} = f_{2}(x_{1},x_{2}) \hspace{.2cm} s.t. \frac{dx_{1}}{dt} = 0 
\end{gathered}
\end{align}
\indent Many interconnected systems such as control problems in adaptive control, numerical algorithms in optimization theory etc are simpler to design when viewed as above. However, any statements on the steady-state behavior and the stability of the two times scale systems (\ref{eq:b22}) do not automatically hold for the original single scale system (\ref{eq:b21}). One standard way to ensure that the properties of (\ref{eq:b22}) extend to (\ref{eq:b21}) is to enforce a sufficient time-scale separation between the two subsystems and then employ the tools of singular perturbation.\\
\indent Under the assumption that $f_{2}(x_{1},.)$ has a finite number of isolated roots $x_2(x_{1})$ $(\phi(x_{1})$ in $x_{2} \pm\phi(x_{1}) = 0)$ one can define the standard singular perturbation conditioned system.
$$\dot x_{1} = f_{1}(x_{1},x_{2})$$
$$\epsilon \dot x_{2} = f_{2}(x_{1},x_{2})$$
where $0\leq\epsilon\leq1$ is a design parameter to enforce the desired level of time scale separation.
\indent In the singular limit $\epsilon \rightarrow0$ becomes a degenerate system by Tikhonov's theorem and reduces to (\ref{eq:b22}).\\
\indent Singular perturbation analysis allows to guarantee that if both systems in (\ref{eq:b22}) are asymptomatically stable, then the conditioned  interconnection is also stable (and has the same equilibria) where $\epsilon$ is below a certain threshold $\bar \epsilon$. This type of conditioning comes at a cost as the second subsystem cannot be made arbitrarily fast in practice, the design choice is $\epsilon <<1$ necessarily slows down the first subsystem and thus limits the convergence rate and deteriorates the performance of the entire interconnection.\\
\indent The solution to the above problem is to aim for time-scale separation in the dynamical system (\ref{eq:b21}) which can be achieved through the principle of normally hyperbolic manifold obtained as a result of the P\&I approach discussed in this article.\\
   $x_{2}^s$ as discussed above now defines the implicit manifold as follows
   $$x_{2}^s = M = \{(x_{1},x_{2}) | x_{2} \pm \phi(x_{1}) = 0\}$$
   Then 
   \begin{align}
   R& = \begin{bmatrix}
       (\nabla\phi)^2 & \pm \nabla \phi \\
       \pm \nabla \phi & 1   
   \end{bmatrix} &=\hspace{.1cm}
   \begin{bmatrix}
       m_{11} & m_{12} \\
       m_{21} & m_{22}
   \end{bmatrix}
   \end{align}
   \\
   and thus with the connection $m_{21}m_{22}^{-1} = -\nabla \phi$, 
   one has $V_{H}$ which is tangent to M, and $V_{V}$ where 
   $$V_{V} = \dot x_{2} \pm \nabla \phi \dot x_{1} = f_{2}(x_{1},x_{2})$$
   Hence, for the dynamical system (\ref{eq:b21}) one has: 
   
       $$ \begin{bmatrix}
        \dot{\hat{x_{1}}}\\
        \dot{\hat{x_{2}}}
    \end{bmatrix} = \begin{bmatrix}
        1 & 0\\
        \pm \nabla_{x_1} \phi & 1
    \end{bmatrix} \begin{bmatrix}
        \dot x_{1}\\
        \dot x_{2}
    \end{bmatrix} = \begin{bmatrix}
        f_{1}(x_{1},x_{2}) \\
       \pm {\nabla}_{x} \phi \dot x_1 + f_{2}(x_{1},x_{2})
    \end{bmatrix}$$
The matrix $\begin{bmatrix}
        1 & 0\\
        \nabla_{x_{1}}\phi & 1
    \end{bmatrix}$ is called the sensitivity conditioning matrix in \cite{picallo2022sensitivity} and with $\nabla_{x} \phi$ designated as $S_{x_{1}}^{x_{2}}(x_{1},x_{2})$ \\
 
Note: If is the original system (\ref{eq:b21}) is stable then the second term of the stabilizing control law $\dot x_{2} = -\nabla_{x_{1}} \phi\dot x_{1} - \alpha M(x_{1},x_{2})$ is not required, and only the stable flow of the off-manifold system trajectory is shaped along the fiber direction. The control action $- \alpha M(x_{1},x_{2})$ which drives the system dynamics along fiber direction ($V_{V}$) is used only when the given system's dynamics is unstable or if a control law not dependent on stable flow of the given system is needed. \\
\indent The following subsection briefly discusses two examples highlighting the P\&I approach to control design using time-scale separation. The first example involves a stable linear dynamical system and the P\&I approach is used only for shaping along the fiber direction, and the second example includes control action and shaping along the fiber direction.

\subsubsection*{B.1: Linear system}
Consider a system of linear equations defined as \cite{gajula2022approximate}:
\begin{equation}\label{eq:b11}
\begin{gathered}
    \dot x_{1} = a_{11}x_{1} + a_{12}x_{2}  \\
\dot x_{2} = a_{21}x_{1} + a_{22}x_{2}  
\end{gathered}
\end{equation}
The system is assumed to be stable and the objective is to shape the off-manifold trajectories along the fiber direction which can be found using the P\&I procedure.
\begin{enumerate}
    \item Choose the target dynamics to be: $ \dot x_1=f(x_1)$
    \item This requires the target manifold to be chosen as $\dot x_2=0$, i.e. $M=\{{\{ x_1,x_2\}\mid a_{21}x_{1} + a_{22}x_{2} =0}\} \implies x_2+(a_{21}/a_{22})x_1=0 \implies x_2+\phi(x_1)=0 $
    \item Calculate the connection term: $\nabla_{x_1} \phi=a_{21}/a_{22} $ 
    \item With respect to this connection, the original vector field is decomposed as:\\
   $(\dot x_1, (a_{21}/ a_{22}) \dot x_1) \oplus (0,  \dot x_2 - (a_{21}/ a_{22}) \dot x_1)$
 \item Let the new states be denoted as $ (\hat x_1, \hat x_2)$. The modified dynamics under the direct sum decomposition are:
     $$\begin{bmatrix}
    \dot {\hat{x_{1}}}\\
    \dot{\hat{x_{2}}}
\end{bmatrix} = \begin{bmatrix}
    1 & 0\\
    \frac{a_{21}}{a_{22}} & 1
\end{bmatrix}\ \begin{bmatrix}
    a_{11} & a_{12}\\
    a_{21} & a_{22}
\end{bmatrix}\begin{bmatrix}
    x_{1}\\
    x_{2}
\end{bmatrix}$$
  \item Simplifying the above equations:   
  \begin{align}\label{eq:b12}
\dot {\hat {x_1}} &= a_{11}x_{1} + a_{12}x_{2} \nonumber \\
\dot {\hat {x_2}} &=\frac{a_{21}a_{11}}{a_{22}}x_{1} +a_{21}x_{1} +\frac{a_{21}a_{12}}{a_{22}}x_{2}+ a_{22}x_{2}  
\end{align}
 \item The equations (\ref{eq:b12}) can be written as:
 \begin{align}\label{eq:b13}
\dot {\hat {x_1}} &= a_{11}x_{1} + a_{12}x_{2} \nonumber \\
\dot {\hat {x_2}} &=\underbrace{\nabla \phi \dot x_1 }_{pathway}+a_{21}x_{1} + a_{22}x_{2}  
\end{align}
\end{enumerate} 
Discussion:
\begin{enumerate}
    \item It can be verified that, based on the desired dynamics,  the manifold $M$ is chosen to be $\dot x_2=0$, therefore only the $\dot x_1$ dynamics lies on $M$. 
    \item The connection term  $\nabla \phi \dot x_1 $ allows for decoupling of the  $x_1, x_2$ dynamics. 
    \item The P\&I method is much simpler in design and the Connection term achieves the same form of the Sensitivity function in Equation (28) of \cite{gajula2022approximate}.
    \item Also, the use of singular perturbation analysis for control design which requires artificially induced time scale separation can be overcome by the P\&I approach (details are further discussed in the next example).
\end{enumerate} 
\subsubsection*{B.2: Cascade Control}
\begin{figure}[h!]
    \centering
    \includegraphics[scale=0.4]{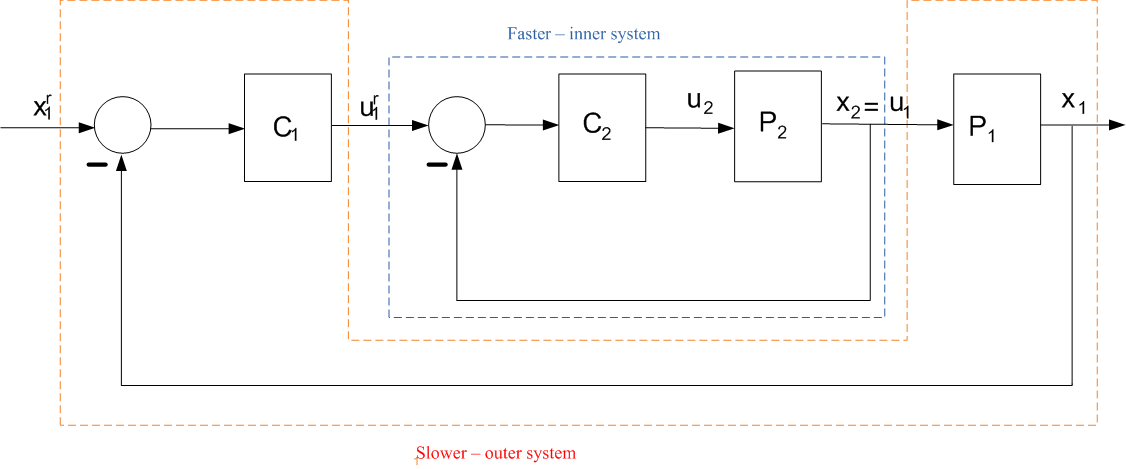}
        \caption{Block diagram of Cascade control}
         \label{fig:cca}
\end{figure}

\begin{figure}[h!]
    \centering
    \includegraphics[scale=0.4]{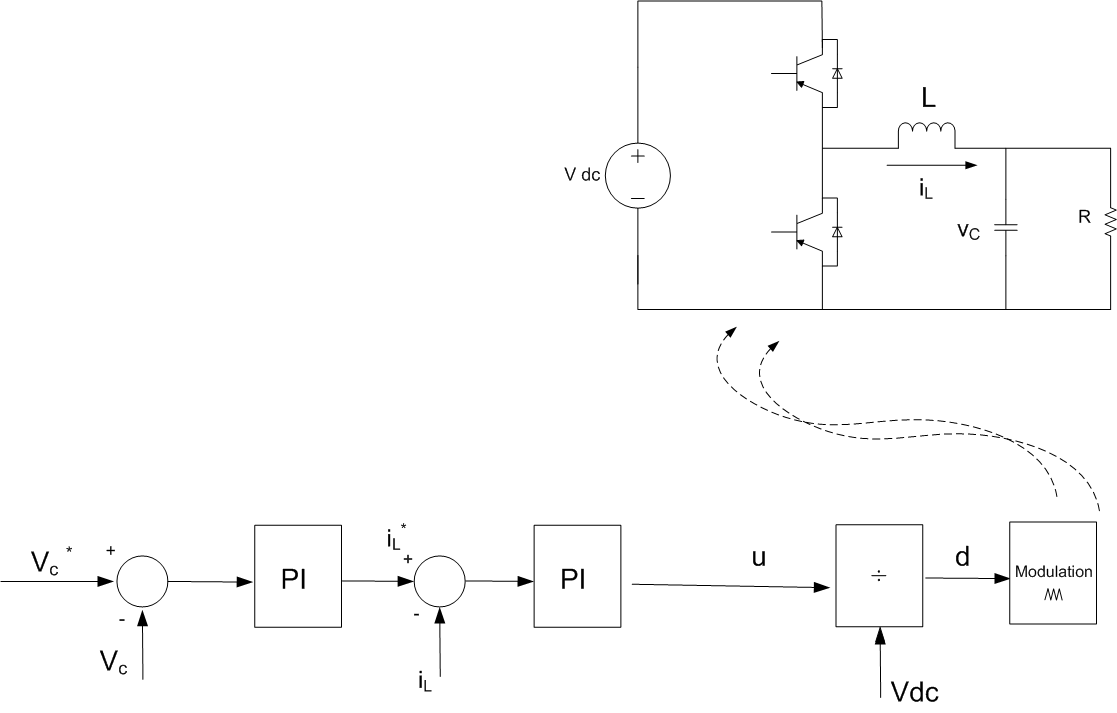}
        \caption{Voltage regulated buck converter with one PI controller}
        \label{fig:dc1}
        
\end{figure}

Consider a standard cascade control architecture  \cite{picallo2022sensitivity} as in Fig \ref{fig:cca}.
It consists of a nested loop, the outer-controlling a primary variable $x_1$ (slower) and the inner loop controlling a secondary variable $x_2$ (faster). The set-point (reference) for the inner loop comes from the outer loop. The aim of this architecture is to ensure that the outer loop looks after the overall control objectives while the inner loop corrects disturbances or responds faster to the inner process variable without affecting the outer loop. However, if the inner loop is slow in responding to the set-point and it receives another set-point, then oscillations are observed in it's transient behavior. In order to avoid this, in general, the inner loop is generally designed to be 3-10 times faster than the outer loop. Yet, the issue of tuning PI for both the controllers $C_1$, $C_2$ so as to obtain the desired objectives still remains challenging.\\
\indent An alternative to address the above issues is to apply the P\&I approach and obtain a time scale separation between the inner loop dynamics and the outer loop dynamics as discussed in the upcoming example. The approach is implemented for a voltage regulated buck converter \cite{gajula2022approximate} with the objective of maintaining a constant output voltage $v_c$  (refer Fig \ref{fig:dc1}). The control strategy involves a cascade control architecture with a decoupled current loop (inner) and a voltage loop (outer) to obtain reference voltage tracking. 
\\
Two cases are considered, one where $u $ is designed using both PI controllers (dual PI cascade controller) and the second where $u$ is designed using the P\&I approach (which requires only one PI controller).  \\
The following notation is adopted: Let $v_c=x_1, \zeta_{v_c}=\zeta_1, i_L=x_2, \zeta_{i_L}=\zeta_2$ where the two additional states for the integral terms: $\zeta_1$ and $\zeta_2$ are defined as:
$$ \dot \zeta_1= (v_c^r-v_c), \dot \zeta_2=(i_L^r-i_L) $$ 
\begin{remark}
In general, the error term is $ x_i-x_i^r$, however, to maintain uniformity with the method used in \cite{picallo2022sensitivity}, \cite{gajula2022approximate}, their notation is adopted.
\end{remark}

Case A: Dual PI Cascade Control
The open-loop state space model for the buck converter is given as:
\begin{align}\label{eq:dcb}
\begin{split}
\begin{bmatrix}
        \dot x_1\\        \dot \zeta_1 \\ \dot x_2\\        \dot \zeta_2
    \end{bmatrix} = \begin{bmatrix}
        -1/RC & 0 & 1/C & 0\\
         -1 & 0 & 0 & 0\\
         -1/L & 0& 0& 0\\
         0 & 0& -1 & 0
        \end{bmatrix} \begin{bmatrix}
        x_{1}\\
        \zeta_1\\
        x_{2}\\
        \zeta_2
    \end{bmatrix} + \\
    \begin{bmatrix}
        0 &0 \\
        1&0\\
        0&0\\
        0&1\\
     \end{bmatrix}\begin{bmatrix}
     v_c^r\\
     i_L^r
     \end{bmatrix}+ \begin{bmatrix}
     0\\0\\1/L\\0 
     \end{bmatrix} u
     \end{split}
\end{align}

Let $ K_{p1}, K_{I1}, K_{p2}, K_{I2}$ be the proportional and derivative gains of the PI controllers of outer and inner loops, respectively. With 
\begin{align} \label{eq:dcb2}
\begin{gathered}
    u=K_{p2}(i_L^r-x_2)+K_{I2}\zeta_2 \\
     i_L^r=K_{p1}(v_c^r-x_1)+K_{p1}\zeta_1
\end{gathered}    
\end{align}
Substitute (\ref{eq:dcb2}) in (\ref{eq:dcb}) :

\begin{align}\nonumber
\begin{split}
\begin{bmatrix}
        \dot x_1\\        \dot \zeta_1 \\ \dot x_2\\        \dot \zeta_2
    \end{bmatrix} = \begin{bmatrix}
        -1/RC & 0 & 1/C & 0\\
         -1 & 0 & 0 & 0\\
         -\frac{1}{L}(1+K_{p1}K_{p2}) & \frac{1}{L}K_{p2}K_{I1}& -\frac{1}{L}K_{p2}& K_{I2}\\
         -K_{p1} & K_{I1}& -1 & 0
        \end{bmatrix} 
    \begin{bmatrix}
        x_{1}\\
        \zeta_1\\
        x_{2}\\
        \zeta_2
    \end{bmatrix}\\ +  
    \begin{bmatrix}
        0  \\
        1\\
       \frac{1}{L}(K_{p1}K_{p2}) \\
        K_{p1}\\
     \end{bmatrix}v_c^r
          \end{split}
\end{align}

Case B: P\&I approach:
Let 
$$ X_1=\begin{bmatrix}
    x_1 \\ \zeta_1
     \end{bmatrix}, 
     X_2=\begin{bmatrix}
    x_2 \\ \zeta_2
     \end{bmatrix}
     $$
\begin{enumerate}
    \item Define the target dynamics: $\dot X_1=f(X_1,\phi(X_1))$ 
    \item Define the manifold $M: \{\{ {X_1, X_2}\} \mid i_L^r-X_2=0\}$ 
    \item $X_2=\phi(X_1)=i_L^r$
    \item Compute $\nabla_{X_1} \phi=[-K_{p1} \hspace{.3cm} K_{I1}]^T$
    \item The control law $u$ for tracking the reference input is given as $\nabla_{X_1} \phi \hspace{0.2cm}\dot X_1-\alpha M$

\item Hence the modified equations are given as:
\begin{align}
\begin{split}
\begin{bmatrix}   \dot {\hat{x_1}}\\        \dot {\hat{\zeta_1}} \\ \dot {\hat{x_2}}\\        \dot {\hat{\zeta_2}}
    \end{bmatrix}=
\begin{bmatrix}\nonumber
        1 & 0 & 0 & 0\\
         0 & 1 & 0 & 0\\
         K_{p1} & -K_{I1}& 0& 0\\
         0 & 0& 0& 1
        \end{bmatrix}   \begin{bmatrix}   \dot x_1\\        \dot \zeta_1 \\ \dot x_2\\        \dot \zeta_2
    \end{bmatrix} +\begin{bmatrix}   0\\       0 \\ -\alpha M\\        0
    \end{bmatrix}
    \end{split}
\end{align}

\begin{align}\nonumber
= 
\begin{split}  
\begin{bmatrix}
        -1/RC & 0 & 1/C & 0\\
         -1 & 0 & 0 & 0\\
        -\alpha K_{p1}+\frac{K_{p1}}{RC}-K_{I1} & \alpha K_{I1} & -\frac{K_{p1}}{C}-\alpha& 0\\
         -K_{p1} & K_{I1}& -1 & 0
        \end{bmatrix} \\ 
    \begin{bmatrix}
        x_{1}\\
        \zeta_1\\
        x_{2}\\
        \zeta_2
    \end{bmatrix} + 
    \begin{bmatrix}
        0  \\
        1\\
       K_{I1}+\alpha K_{p1} \\
        K_{p1}\\
     \end{bmatrix}v_c^r
          \end{split}
\end{align}
\end{enumerate}    

Simulations of the above equations are carried out for the values of parameters \cite{gajula2022approximate} listed in Table \ref{tab:tab1}.
\begin{table}[h]
    \centering
    \begin{tabular}{|c|c|}
    \hline
      R   & 18.6 $\Omega$ \\
      \hline
      L   & 1mH\\
      \hline
      C & 510 $\mu$ F\\
      \hline
      $K_{p1},K_{I1}$ & 1,30\\
      \hline
      $K_{p2},K_{I2}$ & 1,700\\  
      \hline
    \end{tabular}
    \caption \small{DC-DC buck converter parameters and PI gains for case A}
    \label{tab:tab1}
\end{table}
\begin{figure}[h!]
    \centering
    \includegraphics[scale=0.3]{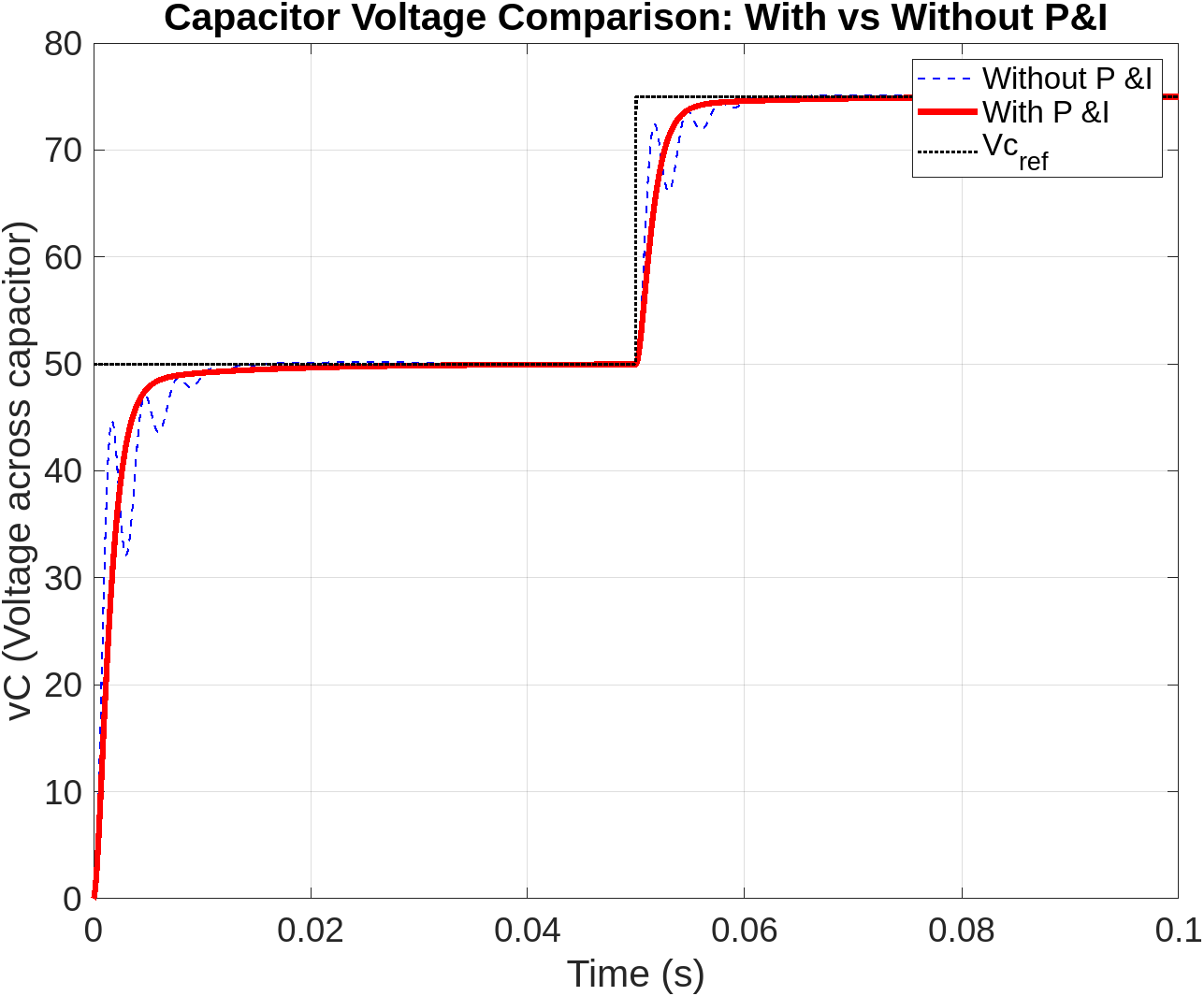}
    \caption{Capacitor voltage $V_c$}
    \label{fig:buckvc}
\end{figure}
\begin{figure}[h!]
    \centering
    \captionsetup{justification=centering}
    \includegraphics[scale=0.3]{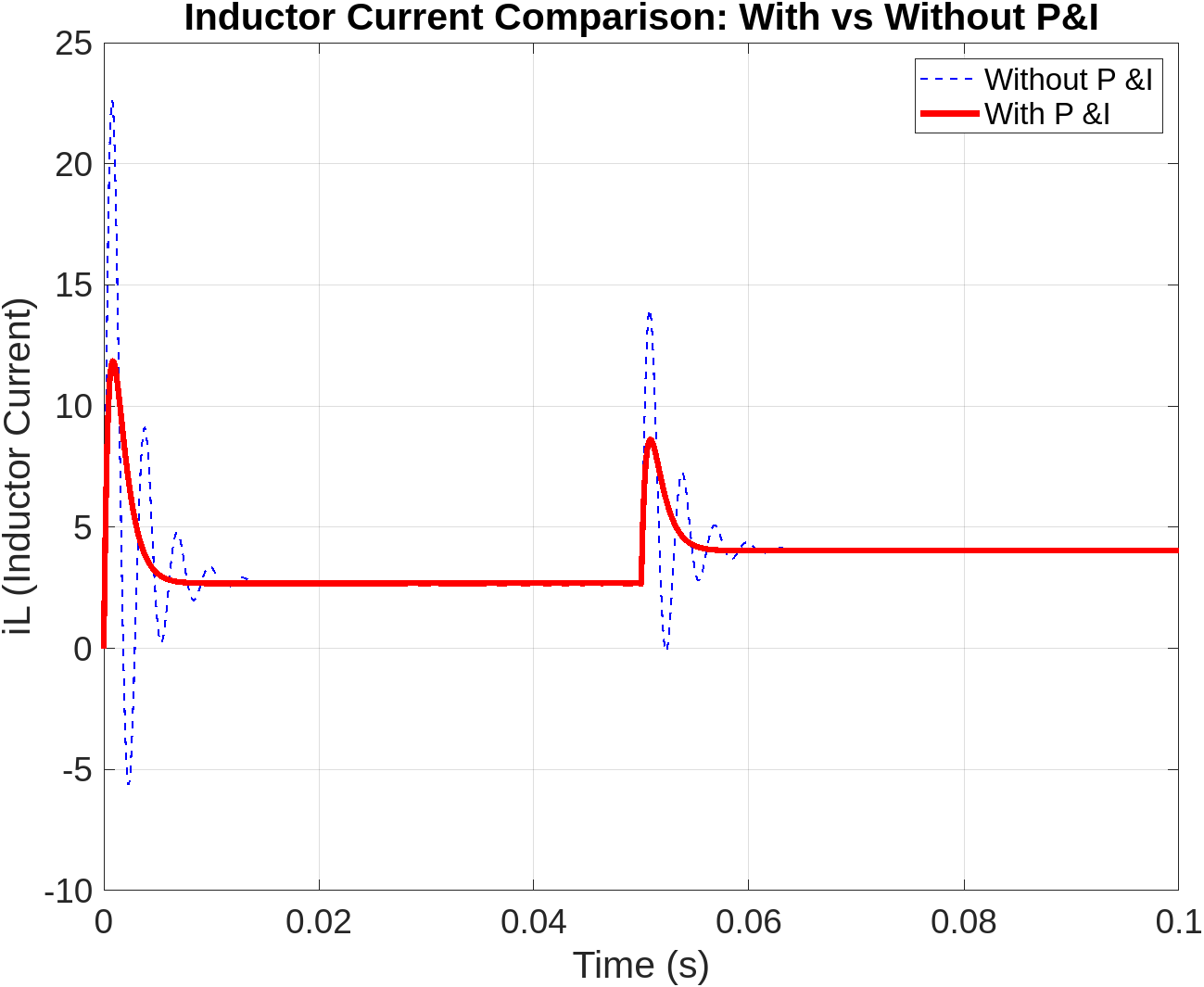}
    \caption{Inductor current $I_L$}
    \label{fig:buckil}
\end{figure}
A minimum steady state error is obtained for $\alpha=700, K_{p1}=1, K_{I2}=100$.
\begin{remark}
The advantages of the P\&I method are listed as follows:
\begin{enumerate}
    \item The basis for obtaining the Sensitivity Function is not clear/ obvious in \cite{picallo2022sensitivity}, \cite{gajula2022approximate} (ideas of the use of Implicit function theorem and a surrogate model for finding $u$). In contrast, the P\&I method proposes defining the manifold $M$ based on the desired dynamics. The connection term $\nabla \phi$ calculated with reference to $M$ renders it to be a normally hyperbolic manifold, with the outer dynamics $(\dot X_1)$ acting tangential to $M$  and the inner dynamics ($ \dot X_2 $) acting orthogonal to $M$. 
    \item The splitting of the tangent space based on $\nabla \phi$ ensures the decoupling of the outer and inner loop dynamics, leading to time-scale separation which is neither based on any ad-hoc method (inner loop should be 3-10 times faster than the outer loop) nor on ideas of singular perturbation theory.
    \item The challenges in tuning of the $K_p, K_I$ values for both the loops (as in the dual PI cascade control) is eliminated, as only a single pair of tuning values are required. The value of $\alpha$ is tuned so that the steady-state error is reduced to zero. This method illustrates how PI tuning can be performed with minimum complexity.
    \item The plots in Fig [\ref{fig:buckvc}], [\ref{fig:buckil}] indicate an improvement in the transient response of $v_C$ and $i_L$, mimicking a behavior which replicates two independent sub-systems.
\end{enumerate}
\end{remark} 

\subsection*{Example C: Parameter Estimation}
Adaptive control systems are generally designed to automatically adjust their parameters in real-time to maintain optimal performance despite changes in system dynamics or external disturbances. In such systems, the use of time-varying input signals is critical for ensuring accurate parameter estimation and system stability. Identifying parameters depends on whether the input signal persistently excites all modes of the system. The modes relate to the dynamic behavior of the system (e.g. damping and frequencies), which are described by the parameters in its mathematical model. Time-varying inputs provide the necessary spectral richness to satisfy Persistent Excitation (PE) conditions. \\ 
\indent A linear regressor equation (LRE) is used to model the relationship between the system's input-output data and its parameters.
 Classically, the Parametric Error Equation (PEE) is formulated which describes the dynamics of the error in parameter estimates. The objective is to solve an optimization problem to drive the estimated error to zero. The PEE turns out to be a linear time-varying (LTV) system. To ensure global exponential convergence of the PEE, the regressor vector must satisfy the persistency of excitation (PE) condition \cite{ortega2020new}. Satisfying the PE property is very rare, hence to alleviate this challenge a relaxed and weaker notion
of Interval excitation (IE) is used in the Concurrent Learning based estimators\cite{5717148}.\\
To achieve faster convergence with the actual parameters, an alternative filtering approach Dynamic Regressor Extension and Mixing (DREM) is proposed in \cite{ortega2020new}. The implementation of a DREM algorithm for $n$ unknown parameters requires $(n-1)$ number of filters that poses challenges such as poor tuning of filter coefficients and learning rate \cite{korotina2020parameter}.\\
\indent The two main factors that affect the performance of the estimation algorithm are the persistence of excitation of the regressor signal and the decoupling of the estimator dynamics. To enhance the information contained in the regressor signal, it is made of the memory regressor, whereby the regressor signal is shaped by a stable first-order low-pass filter. This introduces an additional time scale (inversely proportional to the location of the filter pole) and enhances the transient response of the estimator. The noise rejection is improved through low-pass filtering, and for time varying parameters it provides a tunable parameter to match the rate of parameter variation. Overall, it enriches the regressor with dynamic information from conversion to a second-order system. The P\&I method overcomes the shortcomings of the above-mentioned methods as follows: 
\begin{itemize}
    \item Decoupling of the estimator dynamics: In P\&I method the Connection term and the resulting direct sum decomposition of the tangent space of the error equation (\ref{eq:r}) ensure that the individual error equations associated with the different parameters evolve along orthogonal paths. This is a step towards decoupling the dynamics of $\theta_i$ and $\theta_j$. In comparison with DREM method the dynamics of LRE is decoupled by transforming the original vector estimation problem into a set of scalar regression problems. This method uses filtering to extend the regressor dynamics and a mixing step that effectively diagonalizes the regressor dynamics, as it transforms the extended regressor matrix into a form where the estimation error dynamics for each parameter are governed by independent scalar equation. The mixing process involves the processing of the extended regressor which involves an algebraic transformation which in turn reduces the extended regressor equation into a set of scalar LREs. The filters used on the extension step are crucial for ensuring that the repressor remains PE or satisfies weaker conditions for like IE.
    \item Filter design challenge: DREM performance is highly dependent on the choice of filters used in the regressor extension step. Poorly designed filters can lead to numerical issues such as ill-conditioned regressor metrics requiring careful tuning. The proposed methodology has no such disadvantage. 
    \item Less computational load: Extension of the regressor matrix and the mixing procedure require additional calculations, such as the determinant or adjoint calculation. This can be a drawback in real-time applications or systems with limited processing power. The P\&I approach to parameter estimator uses a single filter to construct a memory regressor extension (MRE) \cite{kreisselmeier2003adaptive} and since decoupling of the estimator dynamics (error) is achieved by exploiting the geometry of the estimator dynamics, the computational load associated with the DREM method does not exist. Thus, the proposed approach to designing the parameter estimator avoids the complications of implementing multiple filters and the choice of higher learning rates. The Gradient Estimator (GE) is cast as a manifold stabilization problem to design a novel framework of Controlled Gradient Estimator (CGE). Better transient performance and noise rejection have been observed in comparison to the DREM procedure.
\end{itemize}
\subsubsection{Parametric Error Equation:}
 Let $y(t) \in  R$ denote the measurable output signal of the system, $\phi(t) \in R^q$ represent the known measurable regressor vector (input vector), $ \epsilon(t)$ be a (generic) exponentially decaying signal and $ \theta \in R^q$ be a constant vector of the unknown
parameters of the system. Then,
\begin{align}\label{eq:lre} y(t)=\Phi^T(t) \theta + \epsilon \end{align}
 Let the online estimate of the parameter $\theta$ be denoted as $\hat \theta$, and the corresponding estimates of the output be $\hat y =\Phi^T \hat \theta$. Denoting the error terms as:
 \begin{align}\nonumber
\tilde{y} &=y-\hat y \\
 &=\Phi^T(t) \theta- \Phi^T(t) \hat \theta =\Phi^T(t)[\theta- \hat \theta] =\Phi^T(t) \tilde \theta   
 \end{align}

 As $ \hat \theta \rightarrow \theta,  \hat y \rightarrow y,$ or, $ \tilde y \rightarrow 0.$ \\This is an equivalent minimization problem: try to find the parameter estimates $\hat \theta$ such that $ y-\hat y$ is minimized. Hence, the objective is to find the minimum of the quadratic error function $ \frac{1}{2} ( y-\hat y)^2$. The choice to accomplish this objective is the Gradient Estimator (GE) technique, which for the unknown parameters in (\ref{eq:lre}) is of the form:
\begin{align}\label{eq:c1}
\dot {\hat \theta}= -\gamma \Phi(y-\Phi^T\hat \theta) \end{align}
 where $ \gamma$ is the adaptation gain. Substituting $y=\Phi^T \theta$ in the above equation and simplifying:
 \begin{align}\label{eq:r}
 \dot {\tilde{ \theta}}= -\gamma \Phi\Phi^T {\tilde{ \theta}} \end{align}
 Equation (\ref{eq:r}) is a linear time varying system of the form: $\dot {\tilde{ \theta}}= A(t) {\tilde{ \theta}}$ and is known as PEE.
\subsubsection{Definitions}
Persistently Excited signal: A bounded vector signal $\Phi(t) \in R^q$ is Persistently Excited (PE) if $ \exists \hspace{.1cm}T>0, \alpha>0:$
 $$\int_{t}^{t+T} \Phi(\tau)\Phi^T(\tau) \geq \alpha I_q \forall t\ge 0 $$
 Interval Excited signal: A bounded vector signal $\Phi(t) \in R^q$ is Interval Excited (IE) on some interval $[t_0, t_1]:$
 
 $$\int_{t_0}^{t_1} \Phi(t)\Phi^T(t) \geq \alpha I_q \forall t\ge 0 $$
 The convergence of the estimated parameters $(\hat \theta )$ to the actual parameters ($\theta$) is greatly dependent on $ A(t)$, which in turn is determined by whether the input signal is PE or IE. If the input signal is PE, the GE technique ensures exponential convergence but exhibits poor transient behavior including oscillations and/or overshoots. If the input signal is IE, then the convergence of the error in the short time window of the input regressor seems challenging, which results in a poor estimate of the actual parameters. 
 
 The P\&I approach proposes to address both the mentioned issues (by choosing a desired dynamics and the corresponding implicit manifold, splitting the original dynamics along $ V_H$ and $V_V$ with respect to the connection term and subsequent calculations), resulting in the improved transient for an input signal which is PE and convergence of (\ref{eq:r}) to equilibrium for an input signal which is IE, as illustrated in the examples below. 
\subsubsection{P\&I approach to solve PEE}
Let $q=2$ (two parameters) and $$ \Phi=[\Phi_1\hspace{0.2cm} \Phi_2]^T$$
The LRE (\ref{eq:lre}) becomes:
$ y=\Phi_1\theta_1+\Phi_2\theta_2$\\
Multiplying throughout by $ \Phi$ and applying a first-order filter (Memory Regressor): $ H(s)=\frac{1}{s+1}$:
$$\frac{1}{s+1} \begin{bmatrix}
    \Phi_{1} y\\ \Phi_{2}y
\end{bmatrix}=\frac{1}{s+1}\begin{bmatrix}
    \Phi_{1} \Phi_{1}& \Phi_{1}\Phi_{2}\\
    \Phi_{2} \Phi_{1}& \Phi_{2}\Phi_{2}
\end{bmatrix}\begin{bmatrix}
    \theta_{1} \\ \theta_{2}
\end{bmatrix}
$$
Resulting in the matrix form:
\begin{equation} \label{eq:mlre} \begin{bmatrix}
    Y_{1} \\ Y_{2}
\end{bmatrix}=\begin{bmatrix}
    \Omega_{11} & \Omega_{12}\\
    \Omega_{21} & \Omega_{22}
\end{bmatrix}
\begin{bmatrix}
    \theta_{1} \\ \theta_{2}
\end{bmatrix}
\end{equation}
Equation (\ref{eq:mlre}) is the modified LRE, resulting in the modified PEE given as: \hspace{0cm}  $ \dot {\tilde \theta} =-\gamma \hspace{0.1cm}\Omega \hspace{0.1cm} {\tilde \theta}$. For $q=2$, the expanded PEE is:
\begin{align}
   \dot {\tilde {\theta_1}} =-\gamma \Omega_{11} {\tilde {\theta_1}}-\gamma \Omega_{12} {\tilde {\theta_2} }\nonumber \\
   \dot {\tilde {\theta_2}} =-\gamma \Omega_{21} {\tilde \theta_1}-\gamma \Omega_{22} {\tilde \theta_2}
\end{align}
Applying the P\&I approach to improve the transient response of the estimator and achieve faster convergence:
\begin{enumerate}
    \item Let the target dynamics be: $\dot {\tilde {\theta_1}}=f( \tilde {\theta_1}, \Phi(\tilde {\theta_1}))$
    \item Define a manifold $M$:
$ \Psi(\tilde \theta)=\{ \{\tilde \theta_1, \tilde \theta_2\} \mid \tilde \theta_2-\beta \tilde \theta_1 =0 \}$
\item Calculate $\nabla \Psi= -\beta$.
\item With reference to the connection $ \nabla \Psi $, the tangent space is split into $V_H$ and $V_V$. Faster convergence of $\tilde \theta_1, \tilde \theta_2$ is achieved when $ \nabla \Psi \dot {\tilde {\theta_1}}$ dynamics is added to $\dot {\tilde {\theta_2}}$ dynamics. 
\item This results in the following modified CGE equations:
\begin{align}\label{eq:c3}
\underset{\dot{\tilde{\theta}}}{
\underbrace{
\begin{bmatrix}
   \dot{\tilde{\theta}}_1 \\
   \dot{\tilde{\theta}}_2
\end{bmatrix}
}
}
= -\gamma  
\underset{P}{
\underbrace{
\begin{bmatrix}
     1 & 0\\ 
     -\beta  & 2 
\end{bmatrix}
}
}
\underset{\Omega}{
\underbrace{
\begin{bmatrix}
     \Omega_{11} & \Omega_{12}\\ 
     \Omega_{21}  & \Omega_{22}
\end{bmatrix}
}
}
\underset{\tilde \theta}{
\underbrace{
\begin{bmatrix}
     {\tilde {\theta_1}}\\
     {\tilde {\theta_2}}
\end{bmatrix}  
}
}
\end{align}
\item The modified CGE equations are written in the general form: $ \dot {\tilde {\theta_1}}= -\gamma \hspace{.1cm} P\hspace{.1cm} \Omega\hspace{.1cm} \tilde{\theta}$
\end{enumerate}
\vspace{.3cm}
\subsubsection{Illustrative simulations} 
\begin{figure}[h!]
    \centering
    \includegraphics[scale=0.6]{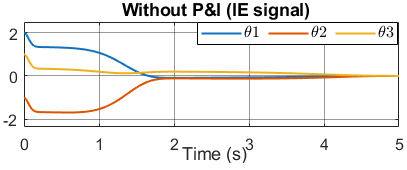}
    \caption{Convergence of errors without P\&I (IE input signal)}
    \label{fig:IE1}
\end{figure}

\begin{figure}[h!]
    \centering
    \includegraphics[scale=0.6]{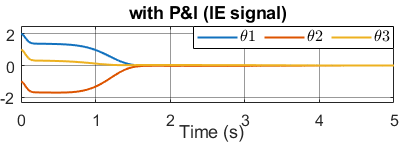}
    \caption{Convergence of errors with P\&I (IE input signal)}
    \label{fig:IE2}
\end{figure}

\begin{figure}[h!]
    \centering
    \includegraphics[scale=0.6]{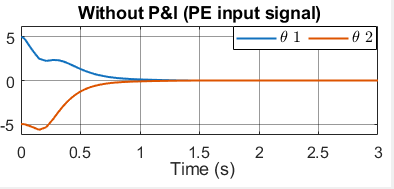}
     \caption{Convergence of errors without P\&I (PE input signal)}
    \label{fig:IE3}
\end{figure}

\begin{figure}[h!]
    \centering
    \includegraphics[scale=0.6]{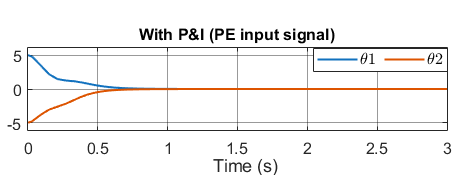}
    \caption{Convergence of errors with P\&I (PE input signal)}
    \label{fig:IE4}
\end{figure}

Consider an IE input signal $\Phi(t)$(q=3):\\ \vspace{0.3cm}
$\Phi(t)=\begin{bmatrix}
    1\\
    cost\\
    \frac{sint +cost}{(1+t)^{0.5}}-\frac{sint}{2(1+t)^{1.5}}
\end{bmatrix}$\\
Simulations of PEE without and with P\&I approach are shown in Fig \ref{fig:IE1} and Fig \ref{fig:IE2} ($\beta=0.9$) respectively. The value of $\gamma=100$ is considered for both cases.\\
\vspace{0.2cm}
Consider a PE input signal $ \Phi(t) =[1 \hspace{0.3cm}sint+sin 3t] $(q=2). Simulation of PEE without and with P\&I approach are shown in Fig \ref{fig:IE3} and Fig \ref{fig:IE4} ($\beta=0.9$) respectively. Values of $\gamma=100$ for both cases.\\
It can be observed from figures that the P\&I approach gives a better transient performance, reductions in oscillations of the error variables and faster convergence for both the PE and the IE input signals.
\section{\label{chap:con}Conclusions}
Fiber bundles provide a powerful framework to model control systems. The advantage is a more general design method with clear separation of local and global behaviors, providing insights into its topological properties, effect of external perturbations, etc., thus enriching the theoretical analysis which aids in developing effective control strategies. 

The proposed P\&I paradigm for control design has found applicability to a wide range of problems associated with shaping the dynamics of a given dynamical system. In addition to the examples discussed in this paper, the method finds usefulness in wider application areas such as incremental backstepping, general non-recursive controller design for higher-order systems, non-triangular non-affine system design, feedforward systems, interlocked systems and orbital stabilization \cite{nayyer2022towards}, synchronization of uncertain chaotic systems, control of systems modeled as Hessian process \cite{NayyerGauss2022}, optimization and control \cite{GunjalCSL2024}, \cite{gunjal2024unified}. 

\section{Appendix}
\subsection{Fiber bundles and Ehresmann Connection} A fiber bundle is a quadruple $(E, \pi, B, F)$ which consists of a total space $E$, a  base space $B$, and fibers $F$ such that $E$ locally resembles the product space $BxF$.
The product space $BxF$ is called a trivial bundle over the base $B$ with fibers $F$. If the fiber of the bundle are homeomorphic to a structure group (a vector space), then the bundle is called a principal bundle (vector bundle). Similarly, once can define an affine bundle, sphere bundle, jet bundle etc. In geometry, Connection provides a way to transport local geometric objects, like tangent vectors, along curves in a consistent manner, allowing for comparisons of local geometries at different points. In the context of fiber bundles, an Ehresmann Connection (EC) defines a parallel transport, essentially "connecting" or identifying fibers over nearby points, which allows for the concept of parallel transport on the bundle. This provides a way to define the curvature of the fiber bundle.  Specifically, in the case of fiber bundles, an EC tells us how movement in the total space induces changes along the fibers. 
\subsection{Connection defined in terms of a form} Consider a bundle $(E, \pi, Q, F)$ with a projection map $\Pi: E\rightarrow Q$ and let $T_q\Pi$ denote its tangent map at any point $q$. We call the kernel of $T_q\Pi$ at any point 'q' the vertical space and denote it by $V_q$. An EC $A$ is a vector valued one-form on $Q$ that satisfies: 
\begin{enumerate}
    \item A is a vector valued:  $A_q:T_q Q\rightarrow V_q$ is a linear map for each point $q\in Q$
    \item A is a projection $A(v_q)=V_q  \forall v_q \in V_q $
\end{enumerate}
If we denote by $H_q$ as the kernel of $A_q$ and call it the horizontal space, the tangent space to $Q$ is the direct sum of $V_q$ and $H_q$, that is, we can split the tangent space $Q$ into horizontal and vertical parts and we can project a tangent vector onto its vertical part using the connection.
The EC creates a direct sum decomposition of the tangent bundle $TE$:
$$TE=H \oplus V $$
The horizontal bundle has the following properties:
\begin{enumerate}
    \item For every point $e \in E$, $H_e $ is a vector subspace of the tangent space $T_qE$ to $E$, called the horizontal subspace of the connection at $q$ and $H_q$ depends smoothly in $q$
    \item The vertical space at Q is tangent to the fibers over $q$
    \item If there is a vertical vector $\forall X_q \in T_qQ$, its horizontal part can be written as:
    $$ hor A_q=X_q-V_q=X_q-A(q)X_q$$
\end{enumerate}
\subsection{Control system represented in fiber bundle framework}
Control of a dynamical system can be briefly expressed as the process of effective manipulation of the given system's vector field to achieve some desired behavior. Therefore, before studying the control objectives, it is important to frame the problem in the right mathematical framework.\\
\indent The theory of fiber bundles provides a geometric framework that has proven to be versatile in tackling the challenges associated with the study of dynamical systems in various scientific domains and control theory in particular \cite{Isidori}, \cite{Schaft}. The topological nature of fiber bundles allows for an exploration of the underlying geometrical properties of control systems, e.g., by examining how the fibers interact with the base space (local and global behavior), one can uncover the topological features that influence controllability and stability. Geometrical insight is crucial for understanding complex systems where the traditional linear approach may fail. In addition, fiber bundles are related to Gauge theory, which plays a significant role in the modeling of complex systems and understanding the effect of external influences on them. In summary, fiber bundles provide a powerful framework for understanding control systems by allowing a sophisticated modeling of the dynamical systems.

\begin{figure}[h!]
    \centering
    \includegraphics[scale=0.65]{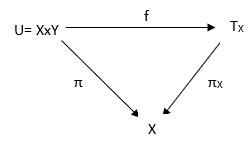}
    \vspace{-0.3cm}
    \caption{Mapping}
    \label{fig:m}
\end{figure}

\indent Let $X$ be a smooth manifold representing the state space of the control system, the total space $U$ of the bundle represents pairs of states and control, i.e., $U=X \times Y$, where $Y$ is another manifold representing the control inputs. There exists a projection map $\Pi: U\rightarrow X$, that associates each point in $U$ (a state-control pair) with the corresponding state in $X$. The above structure allows for a local representation of the control system, where each fiber $U_x=\Pi^{-1} (x)$ consists of all possible control inputs that correspond to a particular state $x \in X$. The dynamics of the control system which depends on the current state and the selected control input can be expressed by a smooth map $f: U \rightarrow T_X$ where $T_X$ is the tangent space of X. A feedback law is defined as a continuous section $u:X \rightarrow U$ such that $\Pi \circ u=idx$. This means that for each state of $X$, there is a corresponding control input in $U$, which allows for the closed loop dynamics described by $f(u(x))$ (Figure \ref{fig:m})
\subsection{Local representation of Connection}
The notion of EC can be locally adapted to a symmetric positive definite (semi-definite) matrix on a manifold $M$ and the connection is compatible with the metric. Consider an $n$-dimensional manifold $M$ with tangent bundle $TM$ such that for an $p\in M$,  $T_pM$ has the following structure:\\
$T_pM=H_p \oplus V_p$ i.e. $H_p \cap V_p=0$\\
The splitting of the tangent space is realized for different local structure of $M$ as follows: 
\paragraph{Case I}
If $M$ is coordinated as $(x,\lambda)$ with $x\in R^k$, $\lambda \in R^{n-k}$ for $k<n$, then $T_pM$ for any $p \in M$ is written as:
$T_pM=(\dot x,0) \oplus (0,\dot \lambda)$
in the Euclidean space with the metric $I_{n \times n}$, i.e.
\begin{equation} \nonumber
\left[ \begin{array}{cc}
\dot x & 0 \\
\end{array} \right]
\left[ \begin{array}{cc}
I_{k \times k} & 0\\
0& I_{n-k \times n-k}
\end{array} \right]
\left[ \begin{array}{cc}
0 \\\dot \lambda \\  
\end{array} \right]=0
\end{equation}
Hence: 
$ (\dot x, \dot \lambda)=(\dot x, 0) \oplus (0, \dot \lambda)
=H \oplus V$
\paragraph{Case II}
Now, if instead of the inner product with respect to the metric 'I', we have the metric $M$, where:
$M_{2 \times 2}=$
$\begin{matrix}
    m_{11} & m_{12}\\
    m_{21} & m_{22}
\end{matrix}$\\
where $ m_{ij} \in R, m_{12}=m_{21}, m_{22}\neq0, |M|\geq0$ \\
The direct sum decomposition (for $\dot x, \dot \lambda) \in R^2$) is:
\begin{equation} \nonumber
 (\dot x, \dot \lambda)=(\dot x, -m_{21}m_{22}^{-1}\dot x) \oplus (0, \dot \lambda+m_{21}m_{22}^{-1}\dot x)   
\end{equation}
 where:
\begin{equation} \nonumber
\left[ \begin{array}{cc}
\dot x & -m_{21}m_{22}^{-1}\dot x \\
\end{array} \right]
\left[ \begin{array}{cc}
m_{11} & m_{12}\\
m_{21} & m_{22}
\end{array} \right]
\left[ \begin{array}{cc}
0 \\\dot \lambda+m_{21}m_{22}^{-1}\dot x \\  
\end{array} \right]=0
\end{equation}
The term $m_{21}m_{22}^{-1}$ is a function of $x$ and can be written as the gradient of any function $q(x)$, that is, $ (m_{21}m_{22}^{-1})=\nabla \phi(x)$.

\paragraph{Case III:}
A general case:
\begin{equation} \nonumber
M=\left[ \begin{array}{cc}
M_{11} \vline & M_{12}\\
\hline
M_{21} \vline & M_{22}
\end{array} \right]=
\left[ \begin{array}{cc|c}
m_{11} & m_{12} & m_{13}\\
m_{21}& m_{22}  & m_{23}\\
\hline
m_{31}& m_{32}  & m_{33}\\
\end{array} \right]
\end{equation}
where $|M|\geq0, m_{22},m_{33}\neq0$.
For $\dot x, \dot y, \dot \lambda \in R^3$, we have the following direct sum decomposition with respect to $M$.
\begin{multline*}
 (\dot x, \dot y, \dot \lambda)=(\dot x, -m_{21}m_{22}^{-1}\dot x,- m_{31}m_{33}^{-1}\dot x -m_{32}m_{33}^{-1}\dot y) \\ \oplus
 (0,\dot y+m_{21}m_{22}^{-1}\dot x, \dot \lambda+m_{31}m_{33}^{-1}\dot x+ m_{32}m_{33}^{-1}\dot y)    
\end{multline*}
   
It can be verified that:
\begin{align} \nonumber
\left[ \begin{array}{cc}
\dot x \\ -m_{21}m_{22}^{-1}\dot x \nonumber \\ -m_{31}m_{33}^{-1}\dot x-m_{32}m_{33}^{-1}\dot y \\
\end{array} \right]^T
\left[ \begin{array}{ccc} \nonumber
m_{11} & m_{12} & m_{13}\\
m_{21}& m_{22}  & m_{23}\\
m_{31}& m_{32}  & m_{33}\\
\end{array} \right] \\
\left[ \begin{array}{cc}\nonumber
0 \\\dot y+m_{21}m_{22}^{-1}\dot x  \\ \dot \lambda+m_{31}m_{33}^{-1}\dot x+ m_{32}m_{33}^{-1}\dot y  
\end{array} \right]= 0
\end{align}
\begin{remark}
The idea of the splitting can be principally extended to the higher dimensions in a rather straightforward manner, and the connection term $m_{21}m_{22}^{-1}$ shall be denoted as the one form $\nabla \phi(x)$ in the rest of the paper.        
\end{remark}
\begin{figure}[!h]
    \centering
    \begin{subfigure}{0.2\textwidth}
        \centering
        \includegraphics[width=1.1\textwidth]{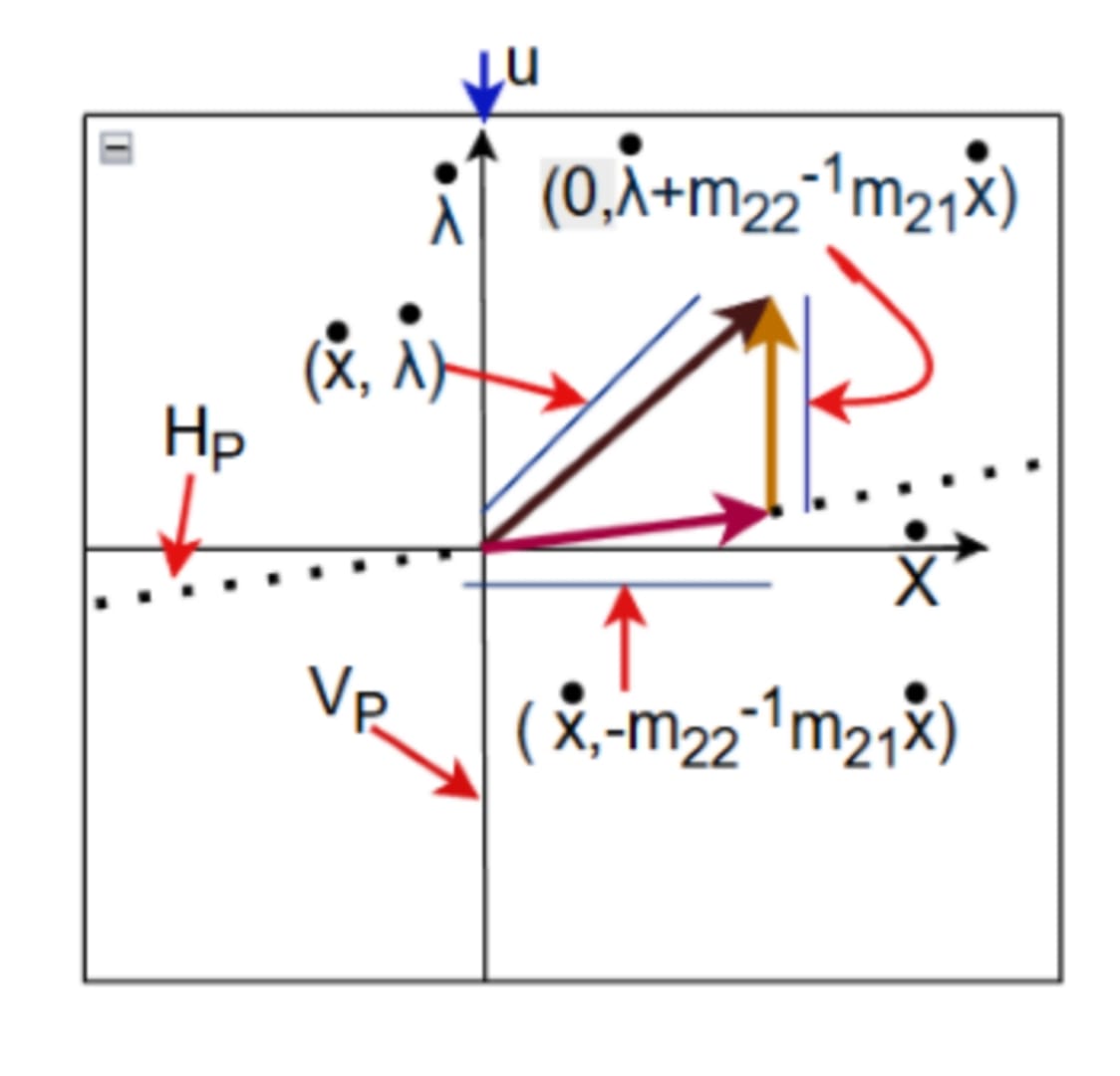}
    \end{subfigure}%
   ~ 
    \begin{subfigure}{0.2\textwidth}
        \centering
        \includegraphics[width=1.105\textwidth]{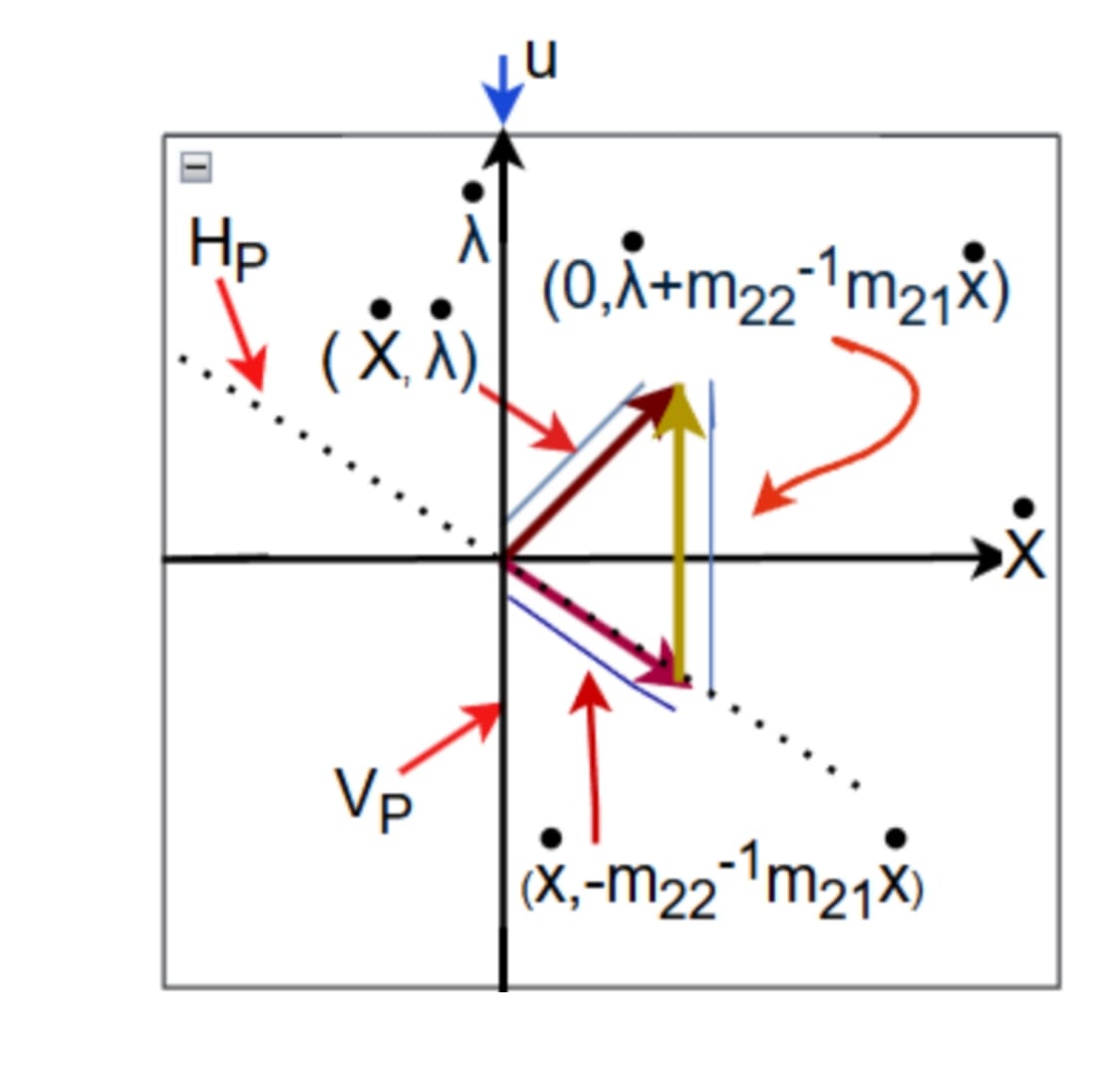}
    \end{subfigure}
    \caption{Geometrical interpretation: vertical vector $V_p$ is along the fiber direction and $H_p \oplus V_p = T_pM$}
    \label{fig:sp}
\end{figure}
\vspace{-0.3cm}
\subsection{Integrability of the connection}
The connection (one form) $\nabla\phi(x)$ is exact.
\begin{corollary}
Integrability of the connection implies that the target manifold $ M $ is obtained by integrating $ V_H$ in the splitting $ T_M=V_H \oplus V_V$, i.e. $V_H$ is tangential to $ M $ everywhere.
\end{corollary}
\textbf{Proof:} For $ \dot \lambda=\mp \hspace{0.1cm} \nabla  \phi(x) \hspace{0.1cm} \dot x,$  i.e.
$ \dot \lambda \pm \hspace{0.1cm} \nabla  \phi(x) \hspace{0.1cm} \dot x=0$. \\
Thus, $M=\int ( \dot \lambda \pm \hspace{0.1cm} \nabla \phi(x) \hspace{0.1cm} \dot x ) dt=\lambda +\phi(x)= c$ \\Hence for $c=0$, the target manifold is defined as :
$$ M=\{ (x, \lambda)| \lambda \pm \phi(x)=0\}$$





\ifCLASSOPTIONcaptionsoff
  \newpage
\fi

\bibliographystyle{IEEEtran}
\bibliography{ref}


\end{document}